\documentclass[aps,pra,twocolumn,showpacs,amsmath,preprintnumbers,nofootinbib,superscriptaddress]{revtex4}

\usepackage{epsfig,verbatim}
\usepackage{amssymb}
 
\usepackage{graphicx}
\usepackage{color}

\def\bfl{\begin{flushleft}}
\def\efl{\end{flushleft}}
\def\bfr{\begin{flushright}}
\def\efr{\end{flushright}}
\def\bc{\begin{center}}
\def\ec{\end{center}}
\def\be{\begin{equation}}
\def\ee{\end{equation}}
\def\ba{\begin{eqnarray}}
\def\ea{\end{eqnarray}}
\def\baa#1{\begin{array}{#1}}
\def\eaa{\end{array}}
\def\bw{\begin{widetext}}
\def\ew{\end{widetext}}
\def\nn{\nonumber }
\def\lb#1{\label{#1}}
\def\bit{\begin{itemize}}
\def\eit{\end{itemize}}

\def\schrod{Schr\"odinger  }

\begin{document}

\preprint{\small Int. J. Mod. Phys. B 27 (2013) 1350163
[arXiv:1207.4877]}

\title{
Non-Hermitian quantum dynamics of a two-level system
and models of dissipative environments
}

\author{Alessandro Sergi}
\email{sergi@ukzn.ac.za}

\author{Konstantin G. Zloshchastiev}
\email{k.g.zloschastiev@gmail.com}

\affiliation{
School of Chemistry and Physics, University of KwaZulu-Natal in Pietermaritzburg,
Private Bag X01, Scottsville 3209, Pietermaritzburg, South Africa\\~~ }

\begin{abstract}
We consider a non-Hermitian Hamiltonian in order to effectively
describe
a two-level system coupled to a generic dissipative environment.
The total Hamiltonian of the model is obtained
by adding a general anti-Hermitian part, depending
on four parameters, to the Hermitian Hamiltonian of a tunneling
two-level system.
The time evolution is formulated and derived in terms of the normalized density operator of the model,
different types of decays are revealed and analyzed.
In particular,  the population difference and coherence
are defined and calculated analytically.
We have been able to mimic various physical situations
with different properties, such as dephasing,
vanishing population difference and purification.
\end{abstract}

\date{\footnotesize Received: 30 May 2012 [JPA], 20 July 2012 [arXiv], 6 June 2013 [IJMPB]}


\maketitle

\section{Introduction}

Non-Hermitian Hamiltonians with complex eigenvalues are finding
numerous applications in
modern physics \cite{nimrod,bender07}. 
Quantum scattering and transport by complex potentials \cite{varga,berg,miro,varga2,muga},
resonances~\cite{nimrod2,seba,spyros}, decaying states \cite{sudarshan},
multiphoton ionization~\cite{selsto,baker,baker2,chu}, optical waveguides~\cite{optics,optics2}
are all examples of the application of non-Hermitian quantum mechanics (NHQM).
It is also known that NHQM can be used in the theory of open quantum systems \cite{wong67,heg93,bas93,ang95,rotter,rotter2,gsz08,bellomo,banerjee,reiter},
one example is given by the Feshbach projection formalism,
which describes a system with a discrete number
of energy levels interacting with a continuum of energy levels~\cite{fesh,fesh2}.

In view of all these applications, the dynamics of non-Hermitian quantum
systems can be regarded as subject of active current research.
It has already been studied by
means of the Schr\"odinger equation~\cite{datto,faisa,baker,baker2}
or derived methods~\cite{thila} while
a different approach has been followed in~\cite{schubert,ghk10},
where the Wigner function
of an initial Gaussian state has been calculated considering correction terms
up to quartic order in the Planck constant.

It is known that a generic open quantum system can be conveniently
described by a density matrix
while wave functions are to be interpreted stochastically \cite{bf}.
In the present work, we consider the density matrix of a two-level system (TLS)
and study its non-Hermitian time evolution
in order to mimic the coupling to a dissipative environment.
The total Hamiltonian of the model has been obtained
by adding a general anti-Hermitian part, depending
on four parameters, to the Hermitian Hamiltonian of a tunneling
two-level system.
The dynamics of the density matrix is solved analytically
in a number of relevant cases.
The parameters are chosen in order to make the evolution
free from singularity and to impose specific constraints
on the nature of the solution.

In section \ref{s-nhd} we give an outline of the NHQM formalism.
In section \ref{s-nhgen} we consider a generic two-level system,
solve the evolution equations and obtain expressions for selected observables.
In sections \ref{s-ecc}, \ref{s-avpdm} and \ref{s-deph} we consider
special cases which correspond to different physical situations,
and
in section \ref{s-mixst} the formalism is applied to mixed states. 
Conclusions are drawn in section \ref{s-con}.


\section{Non-Hermitian dynamics}\lb{s-nhd}

Unless stated otherwise,
the results of this section apply
to a general non-Hermitian system.
Here we will not attempt to give the full
theory of such systems but rather provide the results
which will be most necessary
for what follows.
To begin with, the Hamiltonian operator of any non-Hermitian  system can be  
partitioned into Hermitian and anti-Hermitian parts
\be
\hat{H} = \hat{H}_+ +\hat{H}_- , \lb{e-hplusmin}
\ee
where we denoted $\hat{H}_{\pm}= \pm \hat{H}_{\pm}^{\dag}
= \tfrac{1}{2} (\hat H \pm \hat H^\dag)$.
For further it is convenient to introduce also the self-adjoint operator
$\hat\Gamma \equiv  i \hat H_-$ which
will be referred as the \textit{decay rate operator} throughout the paper.

\subsection{Evolution equations}

Upon introducing the density matrix as 
$\hat{\rho}(t)=|\Psi(t)\rangle\langle\Psi(t)|$,
the non-Hermitian Schr\"odinger equation,
\begin{equation}
\frac{\partial}{\partial t}|\Psi(t)\rangle=-\frac{i}{\hbar}
\left(\hat{H}_++\hat{H}_-\right)|\Psi(t)\rangle\;,
\end{equation}
leads to the evolution of the statistical operator in terms of
a commutator and an anticommutator:
\begin{equation}
\frac{\partial}{\partial t}\hat{\rho}(t)
=-\frac{i}{\hbar}\left[\hat{H}_+,\hat{\rho}(t)\right]
-\frac{i}{\hbar}\left\{\hat{H}_-,\hat{\rho}(t)\right\}\;,
\label{eq:dtrho}
\end{equation}
where the square brackets denote the commutator and the curly brackets
denote the anticommutator, respectively.
From now on we assume that this equation is valid for mixed states as well.
If all the operators are represented by full-rank matrices and 
$\hat{H}_+$ is invertible then,
as shown in \cite{sergi-commthp}, equation (\ref{eq:dtrho}) 
can  also be written in matrix form as
\begin{equation}
\frac{\partial}{\partial t}\hat{\rho}(t)
=
-\frac{i}{\hbar}
\mbox{\boldmath$\rho$}_{\hat{H}_+}^T\cdot
\mbox{\boldmath$\Lambda$}\cdot
\mbox{\boldmath$\rho$}_{\hat{H}_+}\;,
\end{equation}
where we have defined a matrix super-operator
\begin{equation}
\mbox{\boldmath$\Lambda$}=\left[\begin{array}{cc}
0 & 1+ \hat{H}_-(\hat{H}_+)^{-1}\\
-1+(\hat{H}_+)^{-1}\hat{H}_- & 0
\end{array}\right]
\end{equation}
and the column vector
\begin{equation}
\mbox{\boldmath$\rho$}_{\hat{H}_+}
=\left[\begin{array}{c} \hat{\rho}(t) \\ \hat{H}_+\end{array}\right]\;,
\end{equation}
together with the corresponding row vector $\mbox{\boldmath$\rho$}_{\hat{H}_+}^T$.

Since in NHQM the dynamics is not unitary, the trace of the density operator
is not preserved in general:
\be\lb{e-trrhorate}
\frac{\partial}{\partial t}
{\rm tr} \left(\hat{\rho} \right)
=
\frac{2}{i \hbar}
{\rm tr} \left(\hat{\rho} \hat H_-  \right)
,
\ee
hence, the operator 
$|\Psi\rangle\langle\Psi|$
is not a projector.
Following the standard definition of averages 
in 
quantum mechanics,
we introduce a normalized density operator
\be\lb{e-dmatnorm}
\hat\rho' = 
\hat\rho / {\rm tr} \left(\hat{\rho} \right)
\;.
\ee
In terms of the density operator~(\ref{e-dmatnorm})
the quantum average of an observable $\hat{\chi}=\hat{\chi}(0)$ can
be defined as
\begin{equation}
\langle\hat{\chi}\rangle_t
\equiv
{\rm tr}\left(\hat{\rho}' (t)\hat{\chi}(0)\right)
=
{\rm tr}\left(\hat{\rho}(t)\hat{\chi}(0)\right)/ {\rm tr}\left(\hat{\rho}(t)\right)
\;.
\label{eq:spicture}
\end{equation}
Equation~(\ref{eq:spicture}) clearly reduces to the well-known rule
for calculating statistical averages in Hermitian quantum mechanics
in all cases in which ${\rm tr}\left(\hat{\rho}(t)\right)=1$.
Thus, the usage of the normalized density operator ensures the 
probabilistic interpretation of the approach.
Besides, one can check that the evolution equation for $\hat\rho'$,
\begin{equation}
i \hbar\frac{\partial}{\partial t}\hat{\rho}'
=
\left[\hat{H}_+,\hat{\rho}'\right]
+
\left\{\hat{H}_-,\hat{\rho}'\right\}
-2 \,
{\rm tr} \left(\hat{\rho}' \hat H_-  \right)
\hat{\rho}'
,
\label{eq:dtrhoprim}
\end{equation}
is invariant under the ``gauge''
shift 
$\hat H \to \hat H + \epsilon_0 \hat I$ where $\hat I$
is the identity operator and $\epsilon_0$ is an arbitrary complex c-number.
This ensures that it is the difference of energies, rather than their absolute
values, which is a physical observable - as it takes place in the conventional
quantum mechanics.

It is interesting to mention also
that some time ago Gisin,
based on heuristic considerations,
introduced a non-linear equation to effectively account for dissipative 
effects \cite{gisin,gisin2,gisin3}. 
It turns out that the 
Gisin
equation bears a resemblance to 
the one which can be derived
for the normalized density operator
in our approach.
Indeed, 
upon taking the special case 
$\hat{\rho}' = |\Psi'\rangle \langle \Psi' |$ 
(where $\langle \Psi' | \Psi' \rangle = 1$, according to the definition of $\hat{\rho}' $)
we obtain from (\ref{eq:dtrhoprim})
\begin{equation}
i \hbar \frac{\partial}{\partial t}|\Psi'\rangle=
\left(
\hat H_+ 
+ \hat H_-
-
 \langle \hat H_- \rangle
\right) |\Psi'\rangle
,
\end{equation}
where we denoted
$
\langle \hat H_- \rangle \equiv 
\left\langle \Psi'\right|\hat H_- \left|\Psi'\right\rangle
=
\left\langle \Psi'\right| \hat H - \hat H_+ \left|\Psi'\right\rangle
$.
Here the important difference from the 
Gisin equation is the
appearance of the average of the anti-Hermitian part of Hamiltonian -
instead of the average of the total Hamiltonian (or self-adjoint part thereof).
Notice that the non-linear term
$\langle \hat H_- \rangle |\Psi'\rangle$ 
is a functional which 
brings a wavefunction-dependent contribution to 
the Hamiltonian.
This is yet another example of a profound interplay between
the
physics
of open quantum systems and non-linear quantum mechanics:
environment effects are capable of inducing
effective non-linearities in quantum evolution equations
without undermining the conventional quantum postulates
\cite{bf,gisin,gisin2,gisin3,ks87,various1,various2,various3,various4,various5,various6,various7,various8,various9,various10,various11,az11}.

\subsection{Conserved quantities}

The law of change in time of the determinant of the density operator
can be found using the evolution equation and the matrix identity
$\ln\det{\!\hat M}=\text{tr} \ln{\hat M}$. 
Hence, we obtain
\be
\frac{\partial}{\partial t}
\det{\!\hat\rho(t)}
=
- \frac{2}{\hbar}
\det{\!\hat\rho(t)}
\;
\text{tr}\, \hat\Gamma.
\ee
As long as we are working in the \schrod representation,
we can easily integrate the last equation:
\be\lb{e-detsol}
\det{\!\hat\rho (t)}
=
\det{\!\hat\rho (0)}
\,
e^{- \frac{2}{\hbar} t \, \text{tr}\, \hat\Gamma}
.
\ee
Thus, the decay rate operator 
with a positive trace
makes $\det{\!\hat\rho}$ to vanish at large times whereas
the negative-trace one makes the determinant diverge.
If this trace vanishes then we arrive at the special class
of non-Hermitian models for which this determinant is conserved during evolution, 
a particular example of a model from this class can be found in \cite{gkn10}.
Note that the divergence of $\det{\!\hat\rho}$ does not necessarily mean 
the divergence of the determinant of the normalized density operator (\ref{e-dmatnorm}) .

Another probable candidate for a conserved quantity is the purity. 
This notion can be adapted to the non-Hermitian case in the following way.
In the conventional quantum mechanics the analogue of the pure state described
by the density matrix $\hat\rho_p$ would be the state 
$\left|\Psi\right\rangle\!\left\langle \Psi \right|$ obeying the projectivity (idempotency) property
$\left|\Psi\right\rangle\!\left\langle \Psi \right| \left|\Psi\right\rangle\!\left\langle \Psi \right|
= \left\langle \Psi | \Psi\right\rangle \left|\Psi\right\rangle\!\left\langle \Psi \right|$ 
where the norm $\left\langle \Psi | \Psi\right\rangle  \not= 1$ in general.
The analogue of the normalized density operator $\hat\rho_p'$ would thus be the state 
$
\frac{\left|\Psi\right\rangle\!\left\langle \Psi \right| }{
\left\langle \Psi | \Psi\right\rangle
}  $.
In terms of density operators the (generalized) projectivity criterion can be written as
$
(\hat\rho_p)^2 = {\rm tr} (\hat\rho_p) \hat\rho_p
$
or, alternatively, as
$
(\hat\rho_p')^2 = \hat\rho_p'
$.
Therefore, in terms of the normalized density operator (\ref{e-dmatnorm}) the purity can naturally be
defined in a habitual form:
\be\lb{e-pur}
\widetilde{{\cal P}} (\hat\rho) 
\equiv
{\rm tr} \left(\hat{\rho}'^2\right) 
=
{\rm tr} \left(\hat{\rho}^2\right)/
\left( 
{\rm tr} \hat{\rho} 
\right)^2
,
\ee
such that the condition $\widetilde{{\cal P}} (\hat\rho_p)  = 1$ ensures that
the state represented by $\hat\rho_p$ is a (generalized) projector, as in the conventional quantum mechanics.
Using (\ref{eq:dtrho}) and (\ref{e-pur})
we can derive the rate of evolution of the purity:
\be\lb{e-purder}
\frac{\partial}{\partial t}
\widetilde{{\cal P}}
(\hat\rho)
=
\frac{4}{\hbar}
{\cal R} (\hat\rho, \hat\Gamma)
,
\ee
where we denoted
\bw
\be\lb{e-purate}
{\cal R} (\hat\rho, \hat\Gamma)
\equiv
\frac{
\text{tr} (\hat\rho \,\hat\Gamma) \text{tr} (\hat\rho^2)
-
\text{tr} (\hat\rho^2 \hat\Gamma) \text{tr} \hat\rho
}{
(\text{tr} \hat\rho)^3
}
=
\text{tr} (\hat\rho' \,\hat\Gamma) 
\text{tr} (\hat\rho'^2)
-
\text{tr} (\hat\rho'^2 \,\hat\Gamma)
.
\ee
One can see that the purity is conserved under the general non-Hermitian
evolution (\ref{eq:dtrho}) only if the condition 
$
{\cal R} (\hat\rho, \hat\Gamma) \equiv 0
$
is satisfied at all times. Unlike conventional quantum mechanics,
this condition can be state-dependent.

In the case of a two-dimensional Hilbert space ${\cal H}_2$, which is of interest to
the present work, the density matrix and Hamiltonian are represented 
by $2\times 2$ matrices.
Hence, one finds that ${\cal R} (\hat\rho, \hat\Gamma)$ can be further simplified 
and written in a factorized form as
\be
{\cal R} (\hat\rho, \hat\Gamma)|_{\text{rank} = 2}
=
\frac{\det{\!\hat\rho}}{
(\text{tr} \hat\rho)^3
}
\left(
\text{tr} \hat\rho \;
\text{tr}\, \hat\Gamma
-
2 \,\text{tr} (\hat\rho \,\hat\Gamma)
\right)
=
\det{\!\hat\rho'}
\left(
\text{tr}\, \hat\Gamma
-
2 \,\text{tr} (\hat\rho' \,\hat\Gamma)
\right)
,
\ee 
or, using  Eq. (\ref{e-detsol}),
\be\lb{e-ratesol}
{\cal R} (\hat\rho, \hat\Gamma)|_{\text{rank} = 2}
=
\frac{\det{\!(\hat\rho (0))}}{
(\text{tr} \hat\rho)^3
}
\left(
\text{tr} \hat\rho \;
\text{tr}\, \hat\Gamma
-
2 \,\text{tr} (\hat\rho \,\hat\Gamma)
\right)
e^{-\frac{2}{\hbar} t \, \text{tr}\, \hat\Gamma}
,
\ee 
\ew
so that the purity (\ref{e-pur}) is conserved for any state from ${\cal H}_2$
whose initial
density matrix has zero determinant (one usually assumes that ${\rm tr} (\hat{\rho} (0)) = 1$).
In other words, if 
an ${\cal H}_2$ state is initially pure then 
it stays pure during the non-Hermitian evolution,
a specific example to be shown below.

\subsection{Mixed states}

Mixed state is a statistical ensemble of several pure states
which are described by the density matrices 
$\hat\rho^{(i)} = | \Psi^{(i)} \rangle \langle \Psi^{(i)} |$.
The density matrix of a mixed state can be thus written as
the following linear combination
\be\lb{e-mixst}
\hat\rho =
\sum\limits_{i} p_i
\hat\rho^{(i)}
,
\ee
where the coefficients 
must satisfy the normalization condition
\be
\sum\limits_{i} p_i 
\frac{\text{tr} \hat\rho^{(i)}}{
\text{tr} \hat\rho}
=1
,
\ee
which generalizes the one used in the conventional quantum mechanics.
Therefore, one can introduce the time-dependent functions
\be
p'_i (t) \equiv 
p_i 
\frac{\text{tr}\left(\hat\rho^{(i)} (t)\right)}{
\text{tr}\left( \hat\rho (t)\right)}
, \quad 0 \leqslant p'_i (t) \leqslant 1
,
\ee
and
one can also assume that 
$p_i = p'_i (0)$
provided the corresponding density matrices have equal traces at $t=0$.
Then the normalization condition
takes the habitual form
\be
\sum\limits_{i} p'_i (t)
=1,
\ee
and equation (\ref{e-mixst}) can be rewritten as
\be\lb{e-mixst2}
\hat\rho'(t) =
\sum\limits_{i} p'_i(t)\,
\hat\rho'^{(i)}(t)
,
\ee
where the primed operators are defined according to (\ref{e-dmatnorm}).
Specific examples of how one can apply the formalism to mixed states 
are considered in section \ref{s-mixst}.

\section{Non-Hermitian Two-Level System}\lb{s-nhgen}

We consider a system with ground and excited states denoted by
 $|g\rangle$ and $|e\rangle$, respectively.
In terms of the system state projectors the Pauli operators take the standard form~\cite{gerry}:
\begin{eqnarray}
&&
\hat{\sigma}_x=|e\rangle\langle g|+ |g\rangle\langle e| 
=\left(\begin{array}{cc} 0 & ~~1 \\ 1 &~~ 0 \end{array}\right)
, \\&&
\hat{\sigma}_y = i\left( |e\rangle\langle g|-|g\rangle\langle e|\right)
=\left(\begin{array}{cc} 0 & -i\\ i & 0\end{array}\right)
, \\&&
\hat{\sigma}_z= |e\rangle\langle e|-|g\rangle\langle g|
=\left(\begin{array}{cc} 1 & 0\\ 0 & -1\end{array}\right)
.
\end{eqnarray}
The identity matrix $\hat{I}$ is given by the completeness
relation in the two-state space $\hat{I}=|e\rangle\langle e| + |g\rangle\langle g|$.
In this paper we will be mainly interested in such observables
as the \textit{population difference}
\be
\left\langle \hat{\sigma}_z \right\rangle_t
=
\frac{
\hat\rho_{22} (t) - \hat\rho_{11} (t)
}{
\hat\rho_{11} (t) + \hat\rho_{22} (t)
}
,
\ee
and the \textit{coherence}
\be\lb{e-coher}
\left\langle \hat{\sigma}_x \right\rangle_t
=
\frac{
\hat\rho_{12} (t) + \hat\rho_{21} (t)
}{
\hat\rho_{11} (t) + \hat\rho_{22} (t)
}
,
\ee
where 
$\hat\rho_{i j} (t)$
are the $i j$th components of the density matrix.
One can check that during the evolution the spin averages obey the
following identity  
\bw
\be
\left\langle \hat{\sigma}_x \right\rangle_t^2
+
\left\langle \hat{\sigma}_y \right\rangle_t^2
+
\left\langle \hat{\sigma}_z \right\rangle_t^2
=
1 - 4 \det \hat\rho'
=
1 - 4 
\frac{\det{\!\hat\rho (0)}}{(\text{tr} \hat\rho)^2}
e^{-\frac{2}{\hbar} t \, \text{tr}\, \hat\Gamma}
\leqslant 1
,
\ee
\ew
which means that for pure states the averages lie on the Bloch sphere
$\left\langle \hat{\sigma}_x \right\rangle_t^2
+
\left\langle \hat{\sigma}_y \right\rangle_t^2
+
\left\langle \hat{\sigma}_z \right\rangle_t^2
=
1$.

We assume that in absence of any interaction with the environment (closed-system dynamics),
the two-level system is free to make transitions between its two energy levels.
Such a situation is modeled by the Hermitian Hamiltonian
\be
\hat H_+ = - \hbar \Omega \hat{\sigma}_x \;.
\label{eq:tunnspin}
\ee
In order to formulate the open system dynamics of the model,
we introduce a general anti-Hermitian Hamiltonian of the form
\be
\hat H_- = -i \hbar \Omega \left( a_0 \hat{I} + a_1 \hat{\sigma}_x +
a_2 \hat{\sigma}_y + a_3 \hat{\sigma}_z \right)
\label{eq:henv}
\ee
and add it to the Hermitian operator defined in Eq.~(\ref{eq:tunnspin}).
Equations~(\ref{eq:tunnspin}) and~(\ref{eq:henv}) can be
substituted into Eq.~(\ref{e-hplusmin}) in order to obtain a total non-Hermitian Hamiltonian.
Since the physical (normalized) density
operator does not depend on $a_0 $ the latter can
be chosen \textit{ad hoc} - for instance, as to keep $\hat\rho$
finite at all times $0 \leqslant t \leqslant + \infty$.
The choice of the other coefficients $a_i$, $i=1,...,3$,
is made to impose desired constraints on the non-Hermitian dynamics at long time.
To this end, the coefficients of the anti-Hermitian Hamiltonian in Eq.~(\ref{eq:henv})
can be rewritten in a parametrized form as
\be
a_0 = \gamma, \
a_1 =  \gamma \beta , \
a_3 =  W  = \sqrt{(1+\gamma^2)(1-\beta^2)- a_2^2} 
,
\label{eq:param}
\ee
where the square root in $W$ is  defined up to a sign.
The new parameters will be more convenient to use
in what follows since a solution written in their terms
has more physical clarity and conciseness than when using
the original parameters $a_i$.
The $a_i$ coefficients ($i=0,...,3$) are assumed to be real-valued.
This implies that the additional condition
\be\lb{e-realwcond}
(1+\gamma^2)(1-\beta^2) \geqslant a_2^2
\ee
is fulfilled.

The non-Hermitian Hamiltonian studied in this work
is given below
\bw
\ba
\hat{H} = \hat H_+ - i \hat \Gamma 
= - \hbar \Omega \left[ \hat{\sigma}_x +
i \left( 
\gamma \beta \hat{\sigma}_x 
+ a_2 \hat{\sigma}_y + W \hat{\sigma}_z 
+
\gamma \hat{I} 
\right) \right] ,
\label{e-hosc}
\ea
where the coefficients in Eq.~(\ref{eq:param}) have been used.
The decay rate operator operator thus is given by 
\begin{equation}
\hat\Gamma =  \hbar \Omega \left( \gamma \beta \hat{\sigma}_x 
+ a_2 \hat{\sigma}_y + W \hat{\sigma}_z + \gamma \hat{I}
\right)
,
\label{eq:gammaop}
\ee
such that
\be
{\rm tr}\,\hat\Gamma = 2 \hbar \gamma  \Omega
.
\ee

Considering the Hamiltonian (\ref{e-hosc}) and the initial condition
\be\lb{e-inido1}
\hat\rho (0) = |e\rangle\langle e| =
\left(\begin{array}{cc} 1 & ~~0 \\ 0 &~~ 0 \end{array}\right) 
,
\ee
the corresponding equation of motion (\ref{eq:dtrho}) can be solved analytically.
The closed form for the matrix elements of the density matrix is
given by
\ba
&&
\hat\rho_{11} (t) = \frac{1}{2} \frac{e^{-\Gamma t}}{\beta^2 + \gamma^2} \left[
A_{1} \cos{(\omega t)} + B_{1} \sin{(\omega t)} + C_{1} \cosh{(\Gamma t)}
+ D_{1} \sinh{(\Gamma t)} \right] \;,
\\&&
\hat\rho_{12} (t) = \frac{1}{2} \frac{e^{-\Gamma t}}{\beta^2 + \gamma^2} \left[
A_{2} \cos{(\omega t)} + B_{2} \sin{(\omega t)} + C_{2} \cosh{(\Gamma t)}
+ D_{2} \sinh{(\Gamma t)} \right] \;,
\\&&
\hat\rho_{21} (t) = \frac{1}{2} \frac{e^{-\Gamma t}}{\beta^2 + \gamma^2} \left[
A_{3} \cos{(\omega t)} + B_{3} \sin{(\omega t)} + C_{3} \cosh{(\Gamma t)} +
D_{3} \sinh{(\Gamma t)} \right] \;,
\\&& 
\hat\rho_{22} (t) = \frac{1}{2} \frac{e^{-\Gamma t}}{\beta^2 + \gamma^2} \left[
A_{4} \cos{(\omega t)} + B_{4} \sin{(\omega t)} + C_{4} \cosh{(\Gamma t)}
+ D_{4} \sinh{(\Gamma t)} \right] \;,
\ea
where 
\be\lb{e-drc}
\Gamma = 2 \gamma \Omega,
\
\omega = 2 \beta \Omega
\ee
are, respectively, the decay rate coefficient (equal to the trace
of the decay rate operator up to the Planck constant)
and
the tunneling frequency multiplied by the $\beta$ parameter,
and also we have introduced the following coefficients
\ba
&&
A_1 = \beta^2 + \gamma^2 -W^2 ,\ B_1 = 2\beta W,\ C_1=\beta^2 + \gamma^2 +W^2,\  D_1 = 2 \gamma W,
\nn\\&&
A_2  = - C_2 = i W (1-a_2 + i\gamma\beta) ,\ B_2 = -(\gamma + i\beta)(1-a_2 + i\gamma\beta),\   D_2 = i B_2,
\nn\\&&
A_3 = - C_3 = A_2^* ,\ B_3 = B_2^*,\ D_3 = D_2^*,
\nn\\&&
A_4 = -C_4 = -(1-a_2)^2 - \gamma^2 \beta^2 ,\ B_4 =  D_4 = 0
\;.
\nn
\ea
\ew
It is easy to check that the density matrix is still Hermitian
but its trace is not conserved anymore.
Such non-conservation of the trace represents the fact
that the Hermitian (sub)system is coupled to 
environment, 
which 
is mimicked by the anti-Hermitian part of the Hamiltonian,
so that 
the probability can be lost or gained.

If we introduce the coefficients $A_5 = a_2 -1 + \beta^2$ and
$C_5 = 1 - a_2 + \gamma^2$,
the exact evolution of the trace of density matrix is given by
\be
\text{tr}(\hat\rho (t)) =  \frac{e^{-\Gamma t}}{\beta^2 + \gamma^2} \widetilde{T} (t)
,
\label{eq:trrhot}
\ee
where we denoted
$
\widetilde{T} (t) =A_5 \cos{(\omega t)} + \tfrac{1}{2} B_1
\sin{(\omega t)} + C_5 \cosh{(\Gamma t)} + \tfrac{1}{2} D_1
\sinh{(\Gamma t)}
.
$

One can check that the determinant of the density matrix given
by our solution vanishes
at all times 
which confirms the general formula (\ref{e-detsol})
since the initial density operator (\ref{e-inido1}) has zero determinant.
If one also computes the purity (\ref{e-pur})
one obtains that it equals to one at all times,
$
\widetilde{{\cal P}} (\hat\rho (t))
= 1
,
$
which means that the non-Hermitian dynamics in this case
preserves the purity of the initial state of the Hermitian (sub)system.
It can be easily explained by looking at the
formulae (\ref{e-purder}), (\ref{e-ratesol}) and (\ref{e-inido1}).

Using the solution for density operator and definition (\ref{eq:spicture}), 
one can find also the exact time dependence
of the averages of the Pauli operators
\bw
\ba
&&
\langle \hat\sigma_x \rangle_t =
\frac{1}{\widetilde{T} (t)} \left[ A_{6} \left( \cos{(\omega t)} - \cosh{(\Gamma t)} \right)
- \gamma A_5 \sin{(\omega t)} + \beta C_{5} \sinh{(\Gamma t)} \right]
\;,\label{eq:sigmax-ave}
\\&&
\langle \hat\sigma_y \rangle_t = 
\frac{1}{\widetilde{T} (t)} \left[ W (a_2 - 1) \left( \cos{(\omega t)} - \cosh{(\Gamma t)} \right) 
+ \beta C_5 \sin{(\omega t)} - \gamma A_5 \sinh{(\Gamma t)} \right]
\;,\\&&
\langle \hat\sigma_z \rangle_t =
\frac{1}{\widetilde{T} (t)} \left[ A_7 \cos{(\omega t)} + \tfrac{1}{2} B_{1} \sin{(\omega t)} +
C_7 \cosh{(\Gamma t)} + \tfrac{1}{2} D_1 \sinh{(\Gamma t)} \right] \;,
\label{eq:sigmaz-ave}
\ea
where
$ A_6 = - \gamma\beta W $, $A_7 = C_5 - W^2$ and $C_7 = A_5 + W^2$.
One can also determine the asymptotic value ($t\to\infty$) 
of Eqs.~(\ref{eq:sigmax-ave}-\ref{eq:sigmaz-ave}):
\ba
&&
\lim\limits_{t\to +\infty}\langle \hat\sigma_x \rangle_t =
-(\hbar \Omega)^{-1} \lim\limits_{t\to +\infty}\langle \hat H_+ \rangle_t
= \beta ,\nn
\\&&
\lim\limits_{t\to +\infty} \langle \hat\sigma_y \rangle_t = 
\frac{(a_2 -1)(\gamma + W) + \gamma \beta^2} {a_2 -1 - \gamma (\gamma + W)} \;,
\lb{e-asympsig}
\\&&
\lim\limits_{t\to +\infty} \langle \hat\sigma_z \rangle_t = 
\frac{a_2 + \gamma W}{\gamma^2 +1} \;.
\nn
\ea
\ew

As mentioned above, the average of $\hat{\sigma}_z$ expresses the difference
in the probabilities of occupation of the two levels while
the average of $\hat{\sigma}_x$ is related to the phase
of linear superpositions of the two states.
Hence, the behaviour at large times of the two-level system
with Hamiltonian defined in Eq.~(\ref{e-hosc}), 
whose time dependent properties stem from the non-Hermitian
evolution of the density matrix,
can be controlled by the parameters entering the definition
of the decay rate operator in (\ref{eq:gammaop}).
For illustrative purposes,
in Fig.~\ref{f-f0d9} we plot the analytical solutions of $\langle \hat{\sigma}_z\rangle_t$,
$\langle \hat{H}_+ \rangle_t/\hbar\Omega$, and $-\langle \hat{\Gamma} \rangle_t/\hbar\Omega$
as functions of $2\beta\Omega t$ for $\beta=0.9$, $a_2=0.01$ and $\gamma/\beta=0.2, 0.5, 1$.
\begin{figure}[htbt]
\begin{center}\epsfig{figure=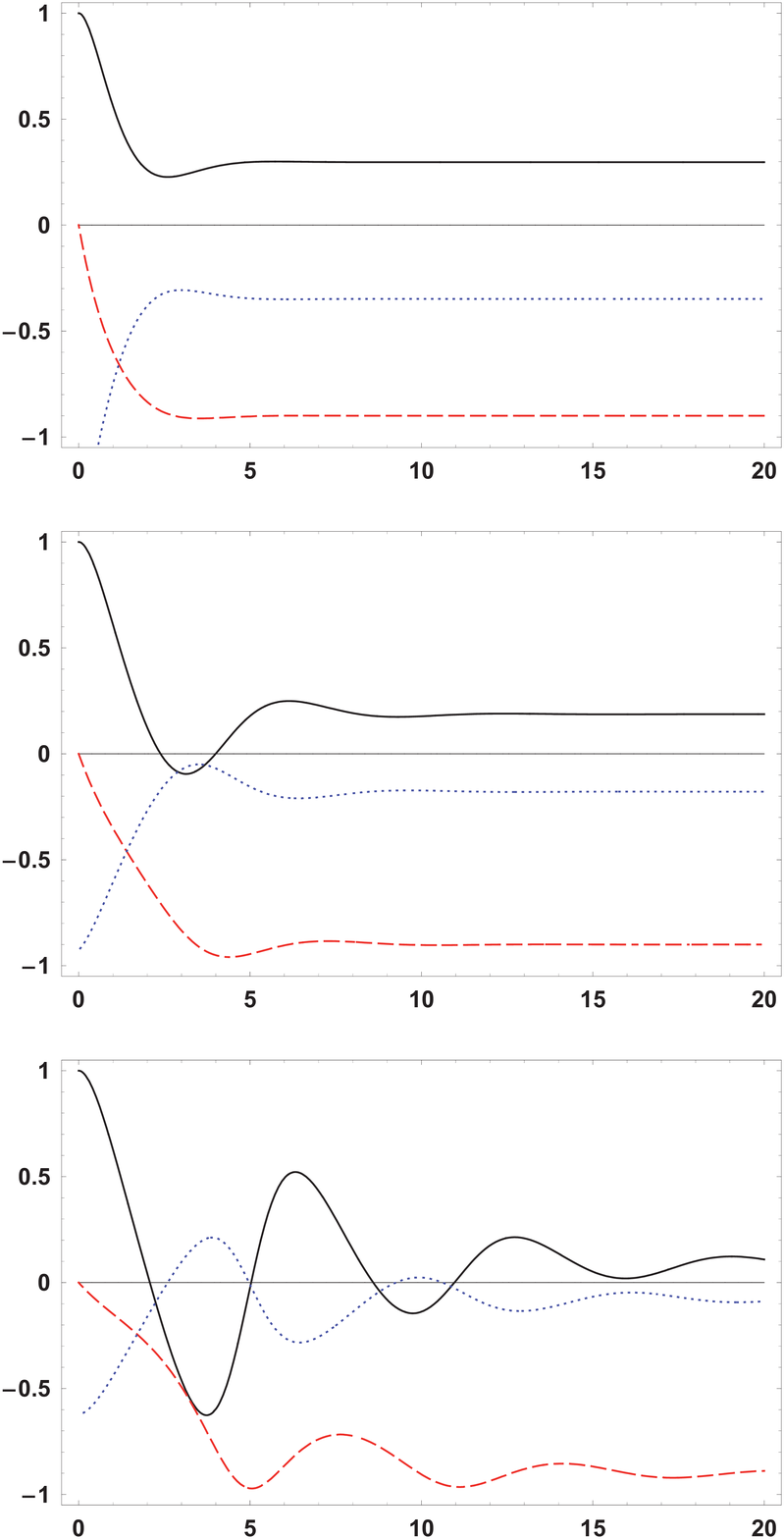,width=  0.95\columnwidth}\end{center}
\caption{Profiles of 
$\left\langle \hat\sigma_z \right\rangle_t $ (solid line),
$ (\hbar \Omega)^{-1} \langle \hat H_+\rangle_t  $ (dashed line),
and $ -(\hbar \Omega)^{-1} \langle \hat \Gamma \rangle_t  $ (dotted line)
versus $2 \beta \Omega t$,
for $\beta = 0.9$, $a_2 = 0.01$  and, from top to bottom, $\gamma /\beta = 1, 0.5, 0.2$.  }
\label{f-f0d9}
\end{figure}

Further, from the constraint in Eq.~(\ref{e-realwcond}) it follows that, 
once $\gamma$ and $a_2$ are fixed,
the oscillatory frequency is always bound from above by the critical value
$\omega_c$.
This latter is, in turn, bound from above by $\omega_+ = 2\Omega $,
i.e.,
$\omega \leqslant \omega_c \leqslant \omega_+$,
with $ \omega_c = \beta_c \,\omega_+$,
$\beta_c^2 = 1 - (a_2^2/(\gamma^2 + 1)) \leqslant \bar\beta^2$,
and $\bar\beta = \pm 1$.
This can be summarized by saying that
at a given decay rate, the parameter $a_2$
measures the difference of the system's critical frequency from the tunneling frequency 
of the Hermitian two-level system whose Hamiltonian is given in (\ref{eq:tunnspin}).


\section{Evolution with conserved average energy}\lb{s-ecc}

There are instances in which the coupling to the environment produces
dissipation while leaving the average energy $\langle \hat H_+ \rangle_t$ 
of the system constant. One such case is provided, for example, by the canonical ensemble. 
In order to describe the relaxation toward the constant average energy condition at large times,
we impose 
\be
\lim\limits_{t\to +\infty}
\langle \hat\sigma_x \rangle_t
=
-(\hbar \Omega)^{-1} \lim\limits_{t\to +\infty}\langle \hat H_+ \rangle_t
= 0
,
\ee
which is equivalent, according to (\ref{e-asympsig}),
to the constraint
\be\lb{e-betcon}
\beta = 0
.
\ee
It turns out that 
this condition leads to the coherence 
(\ref{e-coher}) in this case vanishes not just asymptotically
but identically,
\be
\langle \hat\sigma_x \rangle_t
=
-(\hbar \Omega)^{-1} 
\langle 
\hat H_+ \rangle_t
= 0
,
\ee
which can be shown by directly substituting (\ref{e-betcon}) into (\ref{eq:sigmax-ave}).
This model can be further cast into subclasses,
depending on whether the parameter $a_2$ is chosen to be zero or not.

\begin{figure}[htbt]
\begin{center}\epsfig{figure=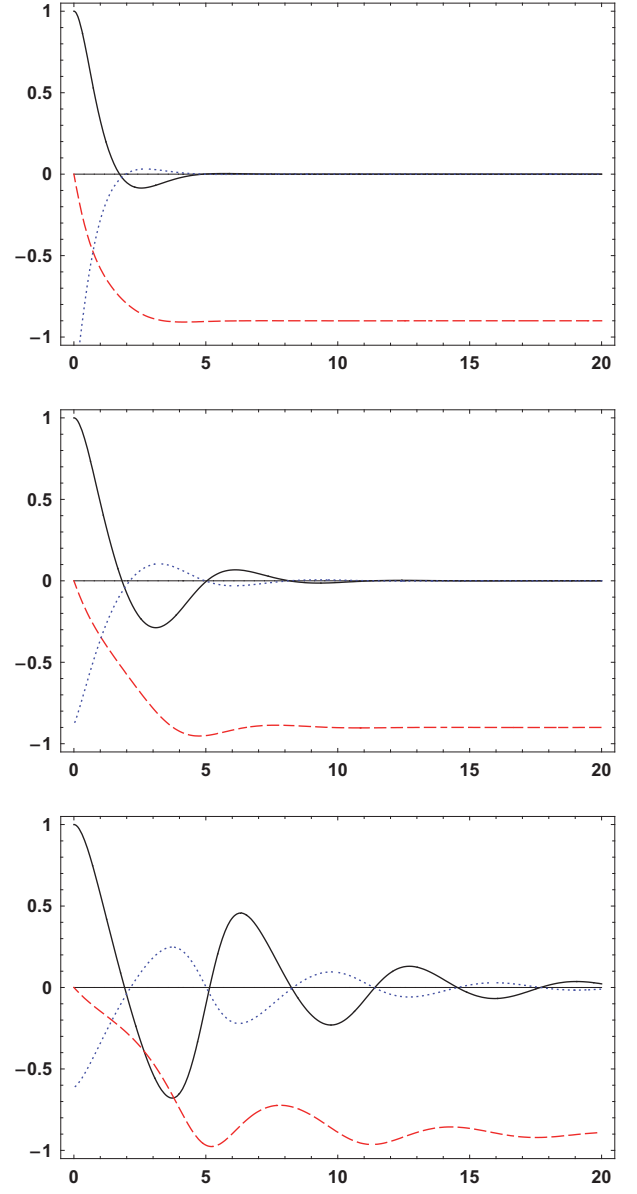,width=  0.95\columnwidth}\end{center}
\caption{Profiles of the averages:
$\left\langle \hat\sigma_z \right\rangle$ (solid line),
$(\hbar \Omega)^{-1} \langle \hat H_+ \rangle_t$ (dashed line)
and 
$- (\hbar \Omega)^{-1}
\langle \hat\Gamma \rangle_t$ (dotted line)
versus time variable $2 \beta \Omega t$,
evaluated at
$\beta = 0.9$,
constraint (\ref{e-a2con})  and, from top to bottom, $\gamma/\beta =$ 1, 1/2, 1/5.
}
\label{f-f0d9sz}
\end{figure}

\subsection{Exponential decay of $\langle \sigma_y \rangle_t$ and $\langle \sigma_z \rangle_t$} 

By this we assume that the observables approach
their asymptotic values exponentially fast.
If we impose 
that $a_2 \not= 0$
then non-Hermitian Hamiltonian is defined by the sum of the Hermitian
Hamiltonian in Eq.~(\ref{eq:tunnspin}) and of the anti-Hermitian decay
operator 
\be
\hat\Gamma^{\text{(ed)}} 
= 
\hbar \Omega \left(a_2 \hat{\sigma}_y
+ \hat{\sigma}_z + \gamma \hat{I} \right)\; .
\ee
multiplied by the imaginary unit.
In such a case, the evolution equation yield the following solution
\bw
\ba
&&
\hat\rho_{11}(t) = \frac{1}{a_2^2} \left[ a_2 \cosh{(\alpha t/2)} -
\sinh{(\alpha t/2)} \right]^2 e^{-\Gamma t} \;,
\\&&
\hat\rho_{12}(t) = -\hat\rho_{21}(t) = \frac{i}{2 a_2^2} (a_2 -1)
\left[ 1- \cosh{(\alpha t)} + a_2 \sinh{(\alpha t)} \right] e^{-\Gamma t} \;,
\\&&
\hat\rho_{22}(t) = \frac{1}{a_2^2} (a_2 -1)^2 \sinh^2{(\alpha t/2)} e^{-\Gamma t}\; ,
\ea
\ew
where $\alpha = 2 a_2 \Omega$ and 
the decay rate coefficient 
$\Gamma$ was defined in (\ref{e-drc}).
If we define
\be
T^{(\text{ed})} (t)
=
(a_2^2 - a_2 +1) \cosh{(\alpha t)} - a_2
\sinh{(\alpha t)} +a_2 -1 \;,
\ee
the trace of density operator can be written as
\be
\text{tr}(\hat\rho (t)) = 
\frac{1}{a_2^2} e^{-\Gamma t} 
T^{(\text{ed})} (t) .
\ee
The analytical expressions of the averages of the spin operators
do not depend on $\gamma$ (yet, their behavior
depends on the sign of $\gamma$ which determines whether the 
exponents are decreasing or increasing with time); they are given below
\ba
&&
\langle \hat\sigma_x \rangle_t = 0 \;,
\\&&
\langle \hat\sigma_y  \rangle_t
=\frac{a_2 -1}{T^{(\text{ed})} (t)}
\left[ \cosh{(\alpha t)}-a_2\sinh{(\alpha t)}-1\right] ,
\\&&
\langle\hat\sigma_z  \rangle_t
=
\frac{a_2}{
T^{(\text{ed})} (t)
}  \left( a_2 - 1 
+ e^{- \alpha t} \right) 
.
\ea

The asymptotic behaviour of the above-mentioned
quantities significantly depends on values of the parameters.
If, for definiteness, one assumes throughout this section that
\be
\gamma > 0
,
\ee
then one can find that
the  parametric space of $a_2$
has the following physically admissible domains:

(i) $0 < a_2 < \gamma$:\\
In this case the density matrix vanishes at large times.
The asymptotic values of other observables
become:
\be
\lim\limits_{t\to +\infty}\langle \hat\sigma_y \rangle_t
= -1
,\
\lim\limits_{t\to +\infty}
\langle \hat\sigma_z \rangle_t
= 
0
.
\ee

(ii) $0 < -a_2 < \gamma$:\\
In this case the density matrix also vanishes at large times,
asymptotic values of other observables
are given by: 
\be
\lim\limits_{t\to +\infty}\langle \hat\sigma_y \rangle_t
= 
\frac{a_2^2 -1}{a_2^2 +1}
,
\ \
\lim\limits_{t\to +\infty}
\langle \hat\sigma_z \rangle_t
= 
\frac{2 a_2}{a_2^2 +1}
.
\ee

\subsection{Polynomial decay of $\langle \sigma_y \rangle_t$ and $\langle \sigma_z \rangle_t$}

By this we assume that the observables approach
their asymptotic values in a way which is described by
a polynomial or rational function of time.
This can be achieved
by choosing the decay operator as
\be
\hat\Gamma^{(\text{pd})}  =  \hbar \Omega 
\left( \hat{\sigma}_z + \gamma \hat{I} \right)\;. 
\ee
The corresponding non-Hermitian Hamiltonian is obtained by adding
(\ref{eq:tunnspin}) to this decay operator (being multiplied by the imaginary unit).
Keeping the initial density matrix in the form of Eq.~(\ref{e-inido1}),
the evolution equation is solved analytically.
The matrix elements have the following explicit time dependence
\ba
&&
\hat\rho_{11}(t) = (\Omega t -1)^2 e^{-\Gamma t} \;,
\\&&
\hat\rho_{12}(t)=-\hat\rho_{21}(t)=i\Omega t\left(\Omega t-1\right)e^{-\Gamma t} \;,
\\&&
\hat\rho_{22}(t) = \Omega^2 t^2 e^{-\Gamma t} \;.
\ea
If we introduce
$
T^{(\text{pd})} (t)
=  2\Omega t (\Omega t -1) +1 
,
$
the trace of density operator is given by
\begin{equation}
\text{tr}(\hat\rho (t)) = e^{-\Gamma t} T^{(\text{pd})} (t) \;.
\end{equation}
The analytical expressions of the averages of the Pauli operators are given
in this case by
\be
\langle \hat\sigma_x \rangle_t = 0
,
\ \
\langle \hat\sigma_y \rangle_t = \frac{2\Omega t (1- \Omega t)}{ T^{(\text{pd})} (t) }
, \ \
\langle \hat\sigma_z \rangle_t = \frac{1- 2\Omega t}{ T^{(\text{pd})} (t) }
,
\ee
and one can see that these observables
evolve according to a polynomial law rather than exponential.
Assuming that $ \gamma > 0 $, the limit for $t\to\infty$ of the density matrix 
and of the averages 
$\langle \hat\sigma_y \rangle_t$
and 
$\langle \hat\sigma_z \rangle_t$ are $-1$ and $0$, respectively.
This is
similar to the case $a_2 > 0$ above,
with
the only difference being that these observables
approach their asymptotic states not exponentially fast
but in a polynomial time.


\section{Asymptotically vanishing population difference}\label{s-avpdm}

For open systems one often expects that the population
difference goes to zero 
at large times:
\be
\lim\limits_{t\to +\infty}
\langle \hat\sigma_z \rangle_t
= 0
.
\ee
This is equivalent, according to (\ref{e-asympsig}),
to the constraints
\be\lb{e-a2con}
a_2 - \gamma \sqrt{1- \beta^2} = 0,
\ \
W  = \sqrt{1- \beta^2}
,
\ee
such that the anti-Hermitian part of the Hamiltonian (\ref{e-hosc}) simplifies to 
\be
\hat\Gamma^{(\text{vp})}
=
-
\hbar \gamma \Omega
\left[
 \beta \hat{\sigma}_x 
-  \sqrt{1- \beta^2}
(\hat{\sigma}_y
+
\gamma^{-1}
\hat{\sigma}_z )
-
\hat{I}
\right]
,
\ee
and
the analytical expressions for a solution from the section \ref{s-nhgen} hold provided one makes
changes in constants according to the constraints (\ref{e-a2con}):
\bw
\ba
&&
\hat\rho_{11} (t) = 
\frac{1}{2} \frac{e^{-\Gamma t}}{\beta^2 + \gamma^2} 
\left[
A_8
\cos{(\omega t)} 
+ \tilde\gamma^2 \cosh{(\Gamma t)}
- 2 W
(\beta \sin{(\omega t)} + \gamma \sinh{(\Gamma t)})
\right] \;,
\\&&
\hat\rho_{12} (t) = 
\frac{1}{2} \frac{e^{-\Gamma t}}{\beta^2 + \gamma^2} 
\left[
A_9
(\cos{(\omega t)} - \cosh{(\Gamma t)})
+ B_5
(\sinh{(\Gamma t)} - i \sin{(\omega t)})
\right] 
,
\\&& 
\hat\rho_{22} (t) = \frac{1}{2} \frac{e^{-\Gamma t}}{\beta^2 + \gamma^2} 
\left[
(2 \gamma W - \tilde\gamma^2)
(\cos{(\omega t)} - \cosh{(\Gamma t)})
\right] 
,
\ea
\ew
where we denoted
\be\lb{e-tgamma}
\tilde{\gamma}=\sqrt{1+\gamma^2}
,
\ee
and
$A_8 = \tilde\gamma^2 - 2 W^2$,
$A_9 = W (\gamma\beta + i (\gamma W -1) )$,
$B_5 = \beta (\tilde\gamma^2 - \gamma W) + i \gamma W (\gamma - W) )$
with $W$ being defined in (\ref{e-a2con}).

The asymptotic values of  observables
become:
\ba
&&
\lim\limits_{t\to +\infty}\langle \hat\sigma_x \rangle_t
=
-(\hbar \Omega)^{-1} \lim\limits_{t\to +\infty}\langle \hat H_+ (t)\rangle
= \beta
,\nn\\&&
\lim\limits_{t\to +\infty}
\langle \hat\sigma_y \rangle_t
= 
-
\sqrt{1- \beta^2},
\ea
whereas
the critical frequency saturates the upper bound,
$\omega_c = \omega_+$ and 
$\beta_c^2 = 1$.
We thus obtain
\be
\beta^2 \leqslant 1 \  \Rightarrow \ 
\omega 
\leqslant \omega_+
,
\ee
the profiles of observables for this model 
are shown in Fig. \ref{f-f0d9sz}.

\section{Dephasing}\label{s-deph}

\begin{figure}[htbt]
\begin{center}\epsfig{figure=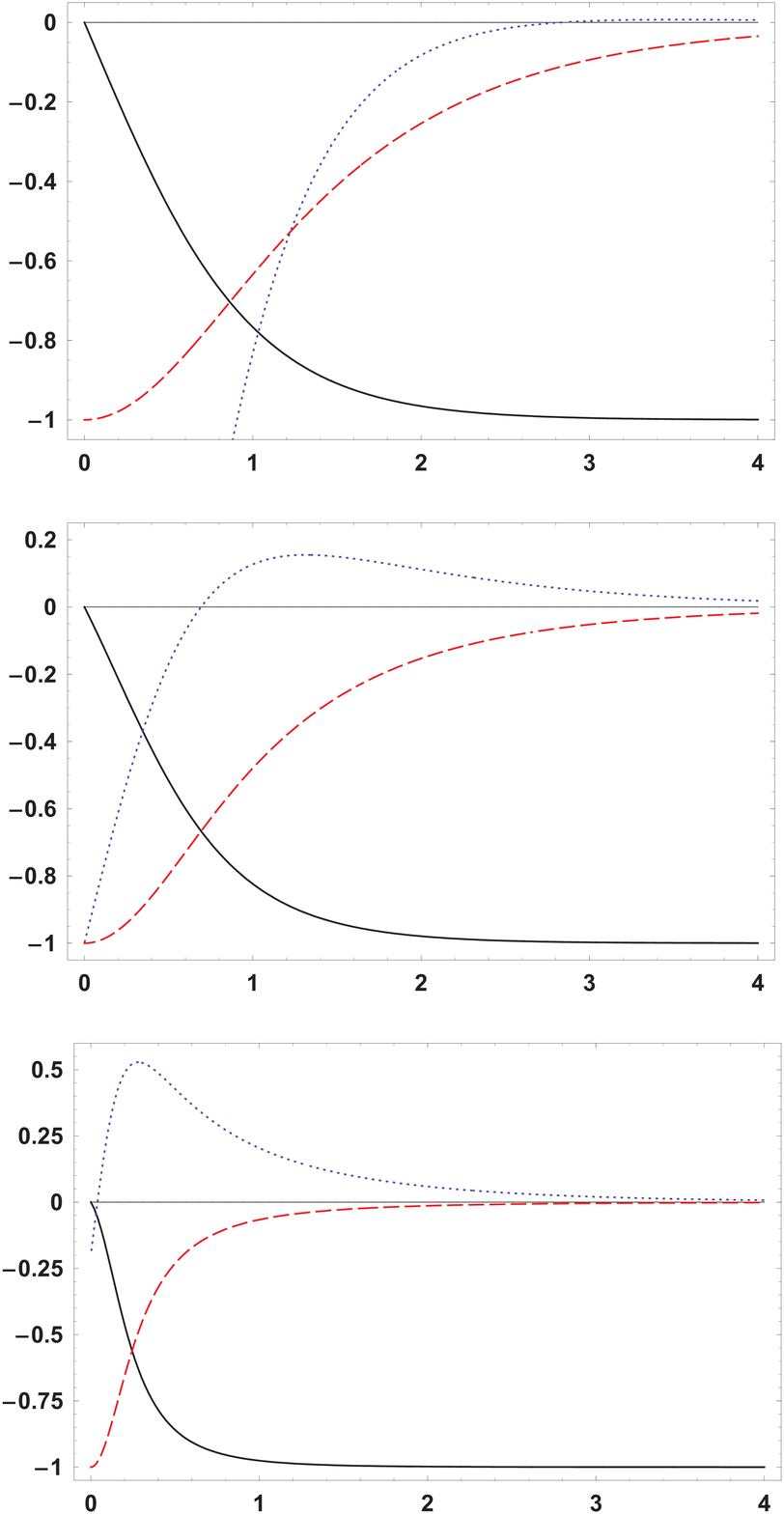,width=  0.95\columnwidth}\end{center}
\caption{Profiles of the averages:
$\left\langle \hat\sigma_z \right\rangle$ (solid line),
$(\hbar \Omega)^{-1} \langle \hat H_+ \rangle_t$ (dashed line)
and 
$-(\hbar \Omega)^{-1}
\langle \hat \Gamma \rangle_t$ (dotted line)
versus time variable $2 \gamma \Omega t$,
evaluated at, from top to bottom, $\gamma =$ 4, 1, 1/5.
}
\label{f-f0dch}
\end{figure}

It can be interesting to investigate whether non-Hermitian dynamics
can cause 
an initial non-diagonal density matrix to become 
diagonal at large times:
this is known as the dephasing.
To this end, let us choose at time zero the following density matrix
\be
\hat\rho (0) = \frac{1}{2}\sum_{k,m=g,e} |k\rangle\langle m|
=
\left(\begin{array}{cc} 1/2 & ~~1/2 \\ 1/2 &~~ 1/2 \end{array}\right) 
.
\ee
If the anti-Hermitian part of the Hamiltonian (\ref{e-hosc}) is chosen 
as
\be
\hat\Gamma^{(\text{dph})} =
-\hbar\Omega
\left[\hat{\sigma}_y-\gamma\left(\sigma_z+\hat{I}\right)
\right]
,
\ee
then the evolution equation yields
\ba
&&
\hat\rho_{11}(t) = \frac{1}{2} e^{- 2 \Gamma t} \;,
\\&&
\hat\rho_{12}(t)=\hat\rho_{21}(t)^* = \frac{1}{2 \gamma} e^{- 2 \Gamma t}
\left[ (\gamma - i) e^{\Gamma t} + i \right]
,
\\&&
\hat\rho_{22}(t)=\frac{1}{2\gamma^2}\left[\left(e^{-\Gamma t}-1\right)^2+\gamma^2\right] 
,
\ea
where the decay rate coefficient was defined in (\ref{e-drc}).

Further, the averages can be written as
\ba
&&
\langle \hat\sigma_x \rangle_t
=
\frac{
\gamma^2
}{
\tilde{\gamma}^2
\cosh{(\Gamma t)}
-
1
}
,\\&&
\langle \hat\sigma_y \rangle_t
=
\frac{
\gamma
(1 - e^{-\Gamma t})
}{
\tilde{\gamma}^2
\cosh{(\Gamma t)}
-
1
}
,\\&&
\langle \hat\sigma_z \rangle_t
=
\frac{
1
-
\cosh{(\Gamma t)}
-
\gamma^2 
\sinh{(\Gamma t)}
}{
\tilde{\gamma}^2
\cosh{(\Gamma t)}
-
1
}
,
\ea  
where
$\tilde\gamma$ was defined in (\ref{e-tgamma}).
We thus obtain the following asymptotics
\ba
&&
\lim\limits_{t\to +\infty}
\hat\rho (t) = 
\frac{1}{2}
\left( 1 + \gamma^{-2} \right) |g\rangle\langle g| \;,
\\&&
\lim\limits_{t\to +\infty}
\langle \hat\sigma_x \rangle_t
=
\lim\limits_{t\to +\infty}
\langle \hat\sigma_y \rangle_t
= 0
,
\\&&
\lim\limits_{t\to +\infty}
\langle \hat\sigma_z \rangle_t
= 
-1
.
\ea
The dephasing is accompanied by the certainty of occupying the ground state
for $t\to\infty$.
The evolution of the observables is shown in Fig. \ref{f-f0dch}.

\section{Mixed states and purification}\label{s-mixst}

The control over the purification of TLS quantum states
is currently of great theoretical \cite{baker90,pur-th,pur-th2,pur-th3}
and experimental \cite{pur-exp,pur-exp2,pur-exp3} interest since
it is important for the quantum state engineering and
quantum computation technology.
In this section we consider the problem 
of the asymptotical purification of mixed states under
non-Hermitian dynamics.
Let us study the  evolution of some 
generic mixed state and find out the conditions under
which it purifies at large times.
As an example, the Hamiltonian is chosen as  a special case of the one given in (\ref{e-hosc}),
\ba
\hat{H} 
= 
- \hbar \Omega \left[ \hat{\sigma}_x +
i \left( 
a_2 \hat{\sigma}_y + W \hat{\sigma}_z 
+
a_0 \hat{I} 
\right) \right] ,
\label{e-hosc-ms}
\ea
where
$W  = \sqrt{1+\gamma^2- a_2^2}$, and $a_0$, $a_2$ and $\gamma$ are free parameters.
The anti-Hermitian part of this Hamiltonian
is not the most general but it is sufficient for
the main purpose of this section.
As in previous sections, the value of $a_0$ does not
enter expressions for the normalized density matrix
and physical observables, and thus it can be left free or it can be chosen \textit{ad hoc}, e.g.,
to make the non-normalized density matrix convergent,
as it happens when one sets $a_0 = \gamma > 0$.
Here we choose the initial condition
\be\lb{e-inido-ms}
\hat\rho (0) = p|g\rangle\langle g| + (1-p) |e\rangle\langle e|
=
\left(\begin{array}{cc} 1-p & ~~0 \\ 0 &~~ p \end{array}\right) 
,
\ee
where $p$ is a constant parameter, $0 < p < 1$.
Solving 
the equation of motion (\ref{eq:dtrho}) with the Hamiltonian (\ref{e-hosc-ms}), we obtain
\bw
\ba
&&
\hat\rho_{11} (t) = 
\Lambda (t) 
\left[
(k_1 (W) - 2 \gamma W) e^{2 \Gamma t}
- 2
(k_1 (\gamma) - 2 \gamma^2) e^{\Gamma t}
+
k_1 (-W) + 2 \gamma W
\right] ,
\nn\\&&
\hat\rho_{12} (t) = 
-
\hat\rho_{21} (t) =
i 
\Lambda (t) 
\left[
\gamma (a_2 + p_1) (e^{2 \Gamma t} -1)
+
W (a_2  p_1 + 1) (e^{ \Gamma t} -1)^2
\right] 
,
\lb{e-sol-ms}\\&& 
\hat\rho_{22} (t) = 
\Lambda (t) 
\left[
k_2 (-W)  e^{2 \Gamma t}
+
2 a_2 (p-1) (e^{\Gamma t} -1)^2
- 2 k_2 (\gamma) e^{\Gamma t}
+
k_2 (W) 
\right]
,\nn
\ea
\ew
and
\be
{\rm tr}\hat\rho (t) = 
2 \Lambda (t) 
\left[
p_+  e^{2 \Gamma t}
+
a_2 p_1 (e^{\Gamma t} -1)^2
- 2  e^{\Gamma t}
+
p_-
\right]
,
\ee
where we denoted
\ba
&&
\Lambda (t)  = (2\gamma)^{-2} e^{- \tilde\Gamma t},
\ 
\tilde\Gamma = \Gamma + 2 a_0 \Omega , 
\nn\\&&
p_1 = 2 p -1, \
p_\pm = \tilde\gamma^2 \pm p_1 \gamma W
,
\nn\\&&
k_1 (\eta) =
\gamma^2 + W^2 + 2 p (1+ a_2 - W^2 + \eta \gamma)
,
\nn\\&&
k_2 (\eta) =
\gamma^2 - W^2 - 2 p (1 - W^2 + \eta \gamma) + 2
,
\nn
\ea
and 
$\Gamma$ and $\tilde\gamma$
have been defined in (\ref{e-drc}) and  (\ref{e-tgamma}), respectively.

The asymptotic behaviour of the density operator crucially depends on
whether the initial-state parameter $p$ lies on the constraint
surface
\be
\Upsilon (p, a_2, \gamma)
\equiv
p_1 (a_2 + \gamma W)
+ \tilde\gamma^2 = 0
.
\ee
We distinguish the following two cases:

(i) Asymptotically pure states.
For those initial states for which 
$
\Upsilon (p, a_2, \gamma)
$
does not vanish at arbitrary $a_2$ and $\gamma$,
the normalized density matrix (\ref{e-dmatnorm}) derived from the solution (\ref{e-sol-ms})
has the following large-times
asymptotics:
\bw
\ba
&&
\hat\rho_{11}' (+\infty) = 
\frac{1}{2 \Upsilon (p, a_2, \gamma)}
\left[
k_1 (W) - 2 \gamma W
\right] ,
\nn \\&&
\hat\rho_{12}' (+\infty) = 
-\hat\rho_{21}' (+\infty)
=
\frac{i}{2 \Upsilon (p, a_2, \gamma)}
\left[
\gamma (a_2 + p_1) 
+
W (a_2 p_1 +  1)  
\right] ,
\\&& 
\hat\rho_{22}' (+\infty) = 
\frac{1}{2 \Upsilon (p, a_2, \gamma)}
\left[
k_2 (-W) + 2 a_2 (p-1)
\right] ,
\nn
\ea
\ew
so one can check that
\be
\lim\limits_{t\to + \infty}\det \hat\rho' (t)= 0
,
\ee
at any non-singular values of the parameters $a_2$ and $\gamma$.

(ii) Asymptotically mixed states.
For those initial states for which $\Upsilon (p, a_2, \gamma)
$
vanishes
(which may happen if the parameters of anti-Hermitian part
obey the inequality 
$
\tilde \gamma^2 <
\left| a_2 + \gamma W \right|
$),
one can derive the condition
for an initial-state parameter:
\be
p=
\frac{1}{2}
\frac{
a_2 + \gamma W - \tilde\gamma^2
}{
a_2 + \gamma W
}
.
\ee 
Under this condition 
the normalized density matrix derived from the solution (\ref{e-sol-ms})
has the following large-times
asymptotics:
\ba
&&
\hat\rho_{11}' (+\infty) = 
\frac{1}{2}
\frac{
(W - \gamma)
(\tilde\gamma^2 + a_2 + \gamma W)
}{
W - a_2 \gamma
}
,
\nn \\&&
\hat\rho_{12}' (+\infty) = 
-\hat\rho_{21}' (+\infty)
=
\frac{1}{2}
i W,
\\&& 
\hat\rho_{22}' (+\infty) = 
\frac{1}{2}
\frac{
(W + \gamma)
(\tilde\gamma^2 - a_2 - \gamma W)
}{
W - a_2 \gamma
} 
,
\nn
\ea
so one obtains
\be
\lim\limits_{t\to + \infty}\det \hat\rho' (t)= 
\frac{\gamma^2}{4} 
\frac{
W^2 (\gamma^2 - 1)
+ 2 a_2 \gamma W
-
(\gamma \tilde\gamma)^2 
}{
(W - a_2 \gamma)^2
} 
,
\ee
which does not vanish
for arbitrary values of the parameters of anti-Hermitian part.

To summarize, for the example considered in this section, 
the purification under non-Hermitian evolution can be controlled by
an appropriate choice of the parameters appearing in the
anti-Hermitian part of the Hamiltonian.

\section{Conclusion}\lb{s-con}

In this paper we have considered a two-level system
described by a Hermitian Hamiltonian and added an
anti-Hermitian term to it.
Hence, we have considered the dynamics resulting
from the total non-Hermitian Hamiltonian.
The anti-Hermitian part
of the Hamiltonian has been assumed to effectively describe  
the averaged influence of degrees of freedom
associated with the environment.
This influence is encoded
in the parameters of the anti-Hermitian part which
can be tuned in order to implement desired properties into a model.
We have also established that the value of the
trace of anti-Hermitian
part plays an important role in determining the decaying
behaviour of the system (or absence thereof).

When analyzing the evolution of the total system,
we have focused on the density matrix of the model
and expressed observables through it.
Analytical solutions have been obtained in a number of relevant cases
- such as when the evolution takes place with dephasing 
or vanishing population difference.
The various types of decays produced show that non-Hermitian dynamics is able
to mimic the effects
of an environment onto a two-level system.
In the case of mixed-state evolution,
we considered as an example a specific Hamiltonian
and clarified the conditions under which
the dynamics led to the purification of the initial state. 
One of the possible future directions would be to apply
this formalism to those physical systems
which allow the two-mode approximation.

\section*{Acknowledgments}

This work is based upon research supported by
the National Research Foundation of South Africa.
The work has been completed during a sabbatical stay
of A.S. at the Department of Physics of the University
of Messina in Italy.
K. Z. is grateful to A. S. for supporting his
visits to the University of KwaZulu-Natal 
and to the University of Messina.





\begin{thebibliography}{}


\bibitem{nimrod}
N. Moiseyev, \textit{Non-Hermitian Quantum Mechanics} 
(Cambridge University Press, Cambridge, 2011).

\bibitem{bender07}
C. M. Bender,
Rep. Prog. Phys. \textbf{70}, 947 (2007).

\bibitem{varga}
B. D. Wibking and K. Varga, Phys. Lett. A {\bf 376} 365 (2012).

\bibitem{berg}
K.-F. Berggreen, I. I. Yakimenko, and J. Hakanen,
New. J. Phys. {\bf 12} 073005 (2010).

\bibitem{miro}
M. Znojil, Phys. Rev. D {\bf 80} 045009 (2009).

\bibitem{varga2}
K. Varga and S. T. Pantelides, Phys. Rev. Lett. {\bf 98} 076804 (2007).

\bibitem{muga}
J.G. Muga, J.P. Palao, B. Navarro, and I.L. Egusquiza,
Phys. Rep. {\bf 395} 357 (2004). 

\bibitem{nimrod2}
N. Moiseyev, Phys. Rep. {\bf 302} 211 (1998).

\bibitem{seba}
W. John, B. Milek, H. Schanz, and P. Seba,
Phys. Rev. Lett. {\bf 67} 1949 (1991).

\bibitem{spyros}
C. A. Nicolaides and S. I. Themelis
Phys. Rev. A {\bf 45} 349 (1992).

\bibitem{sudarshan}
E. C. G. Sudarshan, Phys. Rev. D {\bf 18} 2914 (1978).

\bibitem{selsto}
S. Selst\o, T. Birkeland, S. Kvaal, R. Nepstad, and M. F\o rre,
J. Phys. B: At. Mol. Opt. Phys. {\bf 44} 215003 (2011).

\bibitem{baker}
H. C. Baker, Phys. Rev. Lett. 50, 1579–1582 (1983).

\bibitem{baker2}
H. C. Baker,
Phys. Rev. A {\bf 30} 773 (1984).

\bibitem{chu}
S.-I. Chu and W. P. Reinhardt, Phys. Rev. Lett. {\bf 39} 1195 (1977). 


\bibitem{optics}
C. E. R\"uter, K. G. Makris, R. El-Ganainy, D. N. Christodoulides,
M. Segev, and D. Kip, Nature Physics {\bf 6} 192 (2010).

\bibitem{optics2}
A. Guo, G. J. Salamo, D. Duchesne, R. Morandotti,
M. Volatier-Ravat, V. Aimez, G. A. Siviloglou,
and D. N. Christodoulides,
Phys. Rev. Lett. {\bf 103}  093902 (2009).

\bibitem{wong67}
J. Wong,
J. Math. Phys. \textbf{8}, 2039 (1967).

\bibitem{heg93}
G. C. Hegerfeldt,
Phys. Rev. A \textbf{47}, 449 (1993).

\bibitem{bas93}
S. Baskoutas, A. Jannussis, R. Mignani, and V. Papatheou,
J. Phys. A: Math. Gen. {\bf 26}  L819 (1993).

\bibitem{ang95}
P. Angelopoulou, S. Baskoutas, A. Jannussis, R. Mignani, and V. Papatheou,
Int. J. Mod. Phys. B {\bf 9}  2083 (1995).

\bibitem{rotter}
I. Rotter, arXiv:0711.2926.

\bibitem{rotter2}
I. Rotter,
J. Phys. A {\bf 42} 153001 (2009).

\bibitem{gsz08}
H. B. Geyer, F. G. Scholtz and K. G. Zloshchastiev,
in:
Proceedings of $12^\text{th}$ International Conference on
Mathematical Methods in Electromagnetic Theory
(Odessa, 2008)  250-252.

\bibitem{bellomo}
R. Lo Franco, B. Bellomo, S. Maniscalco, and G. Compagno,
Int. J. Mod. Phys. B {\bf 27} 1345053 (2013).

\bibitem{banerjee}
S. Banerjee and R. Srikanth,
Mod. Phys. Lett. B {\bf 24} 2485 (2010).

\bibitem{reiter}
F. Reiter and A. S. S\o rensen,
Phys. Rev. A {\bf 85} 032111 (2012).

\bibitem{fesh}
H. Feshbach, Ann. Phys. {\bf 5} 357 (1958).

\bibitem{fesh2} 
H. Feshbach, 
Ann. Phys. {\bf 19} 287 (1962).



\bibitem{datto}
G. Dattoli, A. Torre, and R. Mignani, Phys. Rev. A {\bf 42} 1467 (1990).

\bibitem{faisa}
F. H. M. Faisal and J. V. Moloney,
J. Phys. B: At. Mol. Opt. Phys. {\bf 14} 3603 (1981).

\bibitem{thila}
A. Thilagam, J. Chem. Phys. {\bf 136} 065104 (2011).

\bibitem{schubert}
E.-M. Graefe and R. Schubert,
Phys. Rev. A \textbf{83}, 060101 (2011).

\bibitem{ghk10}
E.-M. Graefe, M. H\"oning, and H. J. Korsch,
J. Phys. A {\bf 43} 075306 (2010).

\bibitem{bf}
H.-P. Breuer and F. Petruccione, \textit{The Theory
of Open Quantum Systems} (Oxford University Press, Oxford, 2002).





\bibitem{holyst}
J. A. Holyst and L. A. Turski, Phys. Rev. A {\bf 45} 6180 (1992).

\bibitem{sergi-commthp}
A. Sergi,
Comm. Theor. Phys.  {\bf 56} 96 (2011).

\bibitem{gisin}
N. Gisin, J. Phys. A {\bf 14} 2259 (1981).

\bibitem{gisin2}
N. Gisin,
Physica A \textbf{111}, 364 (1982).

\bibitem{gisin3}
N. Gisin,
J. Math. Phys. \textbf{24}, 1779 (1983).

\bibitem{ks87}
H. J. Korsch and H. Steffen, 
J. Phys. A {\bf 20} 3787 (1987).

\bibitem{various1}
M. D. Kostin, 
J. Chem. Phys. \textbf{57} (1973) 3589.

\bibitem{various2}
M. D. Kostin, 
J. Stat. Phys. \textbf{12} (1975) 145.

\bibitem{various3}
I.~Bialynicki-Birula and J.~Mycielski,
Annals Phys.\  {\bf 100}, 62-93 (1976).

\bibitem{various4}
K.~Yasue,
Annals Phys.\  {\bf 114} (1978) 479.
	
\bibitem{various5}
N. A. Lemos,
Phys. Lett. A \textbf{78} (1980) 239.

\bibitem{various6}
J. D. Brasher,
Int. J. Theor. Phys. \textbf{30} (1991) 979.

\bibitem{various7}
D. Schuch,
Phys. Rev. A \textbf{55}, 935 (1997).

\bibitem{various8}
M. P. Davidson,
Nuov. Cim. B \textbf{116} (2001) 1291.

\bibitem{various9}
J. L. Lopez,
Phys. Rev. E. \textbf{69} (2004) 026110.

\bibitem{various10}
K. G. Zloshchastiev,
Grav. Cosmol. \textbf{16} (2010) 288.

\bibitem{various11}
K. G. Zloshchastiev, Acta Phys. Polon. B \textbf{42} (2011) 261.

\bibitem{az11}
A.~V.~Avdeenkov and K.~G.~Zloshchastiev,
J.\ Phys.\ B: At. Mol. Opt. Phys. {\bf 44} (2011) 195303.


\bibitem{gkn10}
E.-M. Graefe, H. J. Korsch, and A. E. Niederle,
Phys. Rev. A \textbf{82}, 013629 (2010).


\bibitem{gerry}
C. Gerry and P. Knight, \textit{Introductory Quantum Optics}
(Cambridge University Press, Cambridge, 2005).




\bibitem{baker90}
H. C. Baker and R. L. Singleton,
Phys. Rev. A {\bf 42} 10 (1990).


\bibitem{pur-th}
L. P. Hughston, R. Jozsa, and W. K. Wootters, 
Phys. Lett. A \textbf{183}, 14 (1993).

\bibitem{pur-th2}
A. Bassi and G. Ghirardi, 
Phys. Lett. A \textbf{309}, 24 (2003).

\bibitem{pur-th3}
M. Kleinmann, H. Kampermann, T. Meyer, and D. Bruss,
Phys. Rev. A \textbf{73}, 062309 (2006).


\bibitem{pur-exp}
J.-W. Pan, S. Gasparoni, R. Ursin, G. Weihs, and A. Zeilinger,
Nature (London) \textbf{423}, 417 (2003).

\bibitem{pur-exp2}
Z. Zhao, T. Yang, Y.-A. Chen, A.-N. Zhang, and J.-W. Pan, 
Phys. Rev. Lett. \textbf{90}, 207901 (2003).

\bibitem{pur-exp3}
A. E. B. Nielsen, C. A. Muschik, G. Giedke, and K. G. H. Vollbrecht,
Phys. Rev. A \textbf{81}, 043832 (2010).


\end{thebibliography}
\end{document}